# Physical Properties of the Sodium-Based Cubic Fluoro-Perovskites: NaBF$_3$ (B= Ca, Mg or Zn): DFT and TDDFT Studies

S. Idrissi [1], A. Jabar [2,3], H. Labrim [4], L. Bahmad [1,*] and S. Benyoussef [1]

[1] Laboratoire de la Matière Condensée et des Sciences Interdisciplinaires (LaMCScI), Mohammed V University of Rabat, Faculty of Sciences, B.P. 1014 Rabat, Morocco.

[2] LPMAT, Faculty of Sciences Aïn Chock, Hassan II University of Casablanca, B.P. 5366 Casablanca, Morocco

[3] LPHE-MS, Science Faculty, Mohammed V University in Rabat, Morocco

[4] Advanced Systems Engineering Laboratory, National School of Applied Sciences, Ibn Tofail University, Kenitra, Morocco.

Corresponding author: l.bahmad@um5r.ac.ma (L.B)

**Abstract:**[1]

In the present work, we study the structural, electronic and optical properties of Sodium Based Cubic Fluoro-perovskites: NaBF$_3$ (B= Ca, Mg or Zn) using DFT and TDDFT methods. We performed the density functional theory DFT calculations under the mBJ-GGA approximation. In addition, we applied the norm-conserving pseudo-potentials without spin-orbit coupling (SOC) approximations. In addition, we used the Perdew-Burke-Ernzerhof generalized gradient approximation (GGA-PBE) to deduce the physical properties of these alloys. Moreover, the Time-Dependent Density Functional Theory (TDDFT) method has been performed to deduce the optical properties, in an excited state, of the studied Sodium Based Cubic Fluoro-perovskites: NaBF$_3$ (B= Ca, Mg or Zn) materials.

**Keywords:** Fluoro-perovskites; NaBF$_3$ (B= Ca, Mg or Zn); DFT method; TDDFT method.

## I. Introduction

Many researchers have studied the fluoro-perovskites $AMF_3$ material, looking for the mechanisms of the structural phase transitions (SPT) [1-7]. Some other works have pointed out great importance in industrial applications because of their common interesting properties such as structural, electronic, magnetic, optical, elastic and high thermoelectric properties [8-10]. In addition, the unique characteristics of most of the compounds are high-temperature superconductivity, and electric and catalytic behavior [11, 13]. In microelectronics, the telecommunication, the superconductivity, the colossal magneto-resistance and the conductivity are the physical properties of these perovskite materials [14, 15].

Many $AMF_3$ compound types crystallize in the cubic perovskite structure in their highest temperature phase. For lower temperature values, the tilting of 1NT6 octahedral often modifies this structure. The study of such perovskites has been done and an exhaustive description with the classification of octahedral tilting, see Ref. [16]. According to Kassan-Ogly and Naish, the highest-temperature phase is not cubic but pseudo-cubic. The lowest-temperature phase called "ground state" should correspond to static tilts around the three-pseudo cubic axes [17]. Moreover, Flocken et al [5] predicted the ground state of orthorhombic symmetry. From calculations based on ab initio interionic potentials obtained by the Gordon-Kirn approach.

Both $NaMgF_3$ and $NaZnF_3$ such as the family of fluoro-perovskites materials. These materials can undergo phase transitions at high pressure. Recently, two potential candidate structures for the compound $NaMgF_3$, at a very high-pressure value of 450 GaP, were inspected with Pmcn and P63/mmc space groups by Umemoto *et al.* [18]. Nevertheless, the theoretical and experimental studies of Zhao et al. [3] do not prohibit the possibility of the existence of the cubic phase Pm3m. Using X-ray diffraction and Bragg's relation, several authors have provided a critical analysis related to the centrosymmetric distortion of these perovskites. Indeed, they demonstrated two decoupled mechanisms for thermal expansion and compression. Indeed, the thermal expansion is mainly driven by octahedral tilt and the compression is dominated by changes in octahedral bond lengths.

On the other hand, another measure based on the idea cited in Chen et al. [19] explained the existence of an orthorhombic (Pbnm) to cubic (Pm3m) transition when the temperature goes from 900 to 1000 °C. In addition, problems with the use of optical materials for UV and VUV regions present limited transmission. This explains the difficulty in processing and polishing materials due to some materials' cleavage and/or hydroscopic nature. Centro-symmetric materials such as $NaMgF_3$ and $NaZnF_3$, crystallize in cubic perovskite and orthorhombic post-perovskite [20]. Such materials are not faced with these problems and will therefore be suitable for optical applications in the UV and VUV regions [21-24]. In addition, another study was accurately determined and the Pv-pPv phase relationships, in $NaZnF_3$, were examined for the phase transition in materials such as $NaMnF_3$ at high pressure, by Akaogi *et al.* [25]. The cubic phase of the two fluoro-perovskites $NaMgF_3$ and $NaZnF_3$ has been a main topic to obtain a coherent and unified description based on the comparative density functional theory (DFT) of their properties under pressure [26].

In this work, we are investigating structural, electronic and optical properties of Sodium Based Cubic Fluoro-perovskites: $NaBF_3$ (B= Ca, Mg or Zn) using the methods: the Density Functional Theory (DFT), the Time-dependent density-functional theory (TDDFT). In some of our recent works, we have applied the DFT method and other theoretical methods to simulate the physical properties of organic and inorganic Perovskite materials [27-29].

This paper is organized as follows: In section II, we introduce and present the ab-initio method using the DFT approximation. We illustrate the obtained results concerning the structural in sub-section III.1 and the electronic properties of the $NaBF_3$ (B= Ca, Mg or Zn) in sub-section III.2. In section IV, we provide the results of the TDDFT approximation and discuss the obtained results of the studied Fluoro-perovskites: $NaBF_3$ (B= Ca, Mg or Zn). Section V is devoted to the optical properties of the three solar fluoro-perovskites $NaBF_3$ (B= Ca, Mg or Zn). Section VI is dedicated to conclusions.

## II. Computational methodology

According to the density functional theory, we provide a study of the structure, electronic and magnetic properties of the Cubic Fluoro-perovskites: $NaBF_3$ (B= Ca, Mg or Zn) materials, we performed the density functional theory calculations under the quantum Espresso code [30]. We applied norm-conserving pseudopotentials [31] without spin-orbit coupling (SOC) calculations. On the other hand, the band structure was described using the modified Becke-Johnson (mBJ-GGA) potential approximation. The exchange and correlation function used in this work is the Perdew-Burke-Ernzerhof generalized gradient approximation (GGA-PBE) [32, 33].

The configurations of valence electrons for NaCaF$_3$, NaMgF$_3$ and NaZnF$_3$ compounds are the element F has $2s^2$, $2p^5$, while the element Na has $2s^2$, $2p^6$, $3s^1$, the element Ca has $3s^2$, $3p6$, $4s^2$ and Mg has $3s^2$ or the element Zn has $3d^{10}4s^2$.

On the other hand, estimation of the structure and electronic properties of the single cells of solar materials in the cubic crystal phase. The Brillouin area integral was represented by the Monkhorst-Pack k-point scheme, where k points (11×11×11) are used for the first Brillouin area sampling. To optimize the crystal structure and find the ground state energy, Broyden–Fletcher–Goldfarb–Shanno (BFGS) [34] minimization technique is applied. The cut-off energy value has been predicted based on the separation of valence and core energy configuration. For the material studied in this way, this value has been estimated to be about 408 eV. According to the existing literature, all of the forces applied on the atoms converged during geometrical optimization, to a value of 0.002eV/Å. All the properties of the material are calculated after energy optimization. Other studies used this method, see Refs. [35-40].

### III. Results of the DFT Method of the perovskite materials

### III.1. Structural properties

From the existing literature, a large part of Fluoride Perovskites materials has a cubic structure at room temperature [41] of the highest symmetry with the space group of Pm-3m. For these studied NaBF$_3$ (B= Ca, Mg or Zn) compounds, there are two types of positions to build the unit cell. The idea is to take, as the origin atom in position for the first case, either the cation (Na or that of Ca, Mg or Zn) for the second case. The settlement for Ca/Mg/Zn and Na atoms is (0, 0, 0) and (1/2, 1/2, 1/2) respectively, however, the three F atoms are at (1/2, 1/2, 0), (1/2, 0, 1/2), (0, 1/2, 1/2). The Cubic Inorganic Halide Perovskite Materials unit cell is shown in Fig.1 for NaBF$_3$ (B= Ca, Mg or Zn) compounds. Such a figure has been plotted using the Vesta software [42].

In Fig. 2, we provide the optimization of the total energy, of the three fluoro-perovskites NaBF$_3$ (B= Ca, Mg or Zn), as a function of the lattice parameter a (Å). The results presented in this figure have been conducted using the GGA-PBE approximation. It has been found that the most stable material is the NaZnF$_3$, with a lattice parameter value a =4.02 Å. While the less stable fluoro-perovskite is the NaCaF$_3$ compound. In connection with Figure 2, we illustrate in Table 1, a comparison between the results obtained in the present work and the other existing literature concerning the optimal lattice parameter a (Å) of the three studied Fluoro-perovskites NaBF$_3$ (B= Ca, Mg or Zn). The values we found for NaCaF$_3$ and NaZnF$_3$ are noticeably low when compared with the existing literature. While, the lattice parameter value a =3.97 Å, found in this work, for the Fluoro-perovskites NaMgF$_3$, is very close to values provided in the literature, see Table 1.

To determine the most stable structure, the energy of formation was calculated using the following equation

$$E_{form} = E_{tot} - \sum_{x} E_{tot}(x) \qquad (1)$$

$E_{tot}$ is the total energy of the crystal and $E_{tot}(x)$ is the total energy of the mass x with x in (Na, B, F) for NaBF$_3$ with B (Ca, Mg and Zn). The energy of formation is -1.015894 eV, -0.989378 eV and -0.1707 eV respectively for NaCaF$_3$, NaMgF$_3$, and NaZnF$_3$, so we may safely conclude that the system is stable, because $E_{form}$ is negative.

*Table 1: Comparison between the present work and the existing literature for the lattice parameter a (Å) of the Cubic Fluoro-perovskites NaBF$_3$ (B= Ca, Mg or Zn) materials.*

|  | **NaCaF$_3$** | **NaMgF$_3$** | **NaZnF$_3$** |
|---|---|---|---|
| present study | 4.03 Å | 3.97 Å | 4.02 Å |
| existing literature | 4.154 Å [43]<br>4.460 Å [44]<br>4.438 Å [45]<br>4.316 Å [46]<br>4.350 Å [47] | 3.964 Å [48]<br>3.960 Å [49] | 4.065 Å [48]<br>4.041 Å [50]<br>4.000 Å [51]<br>4.065 Å [51] |

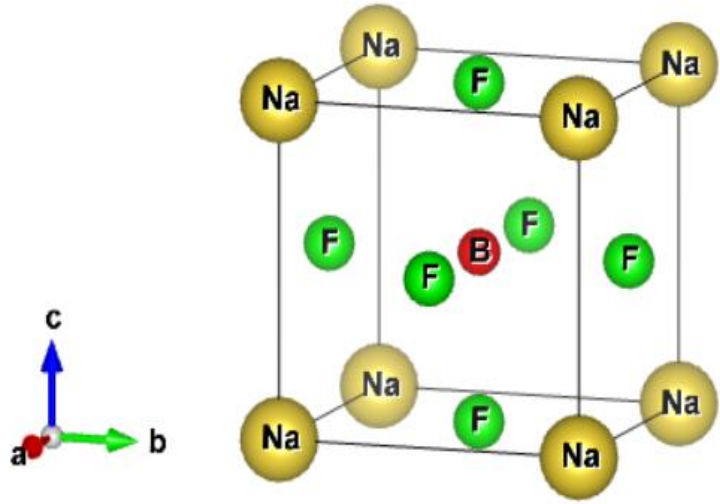

*Fig. 1: The geometry of the studied Fluoro-perovskites NaBF$_3$ (B= Ca, Mg or Zn) materials*

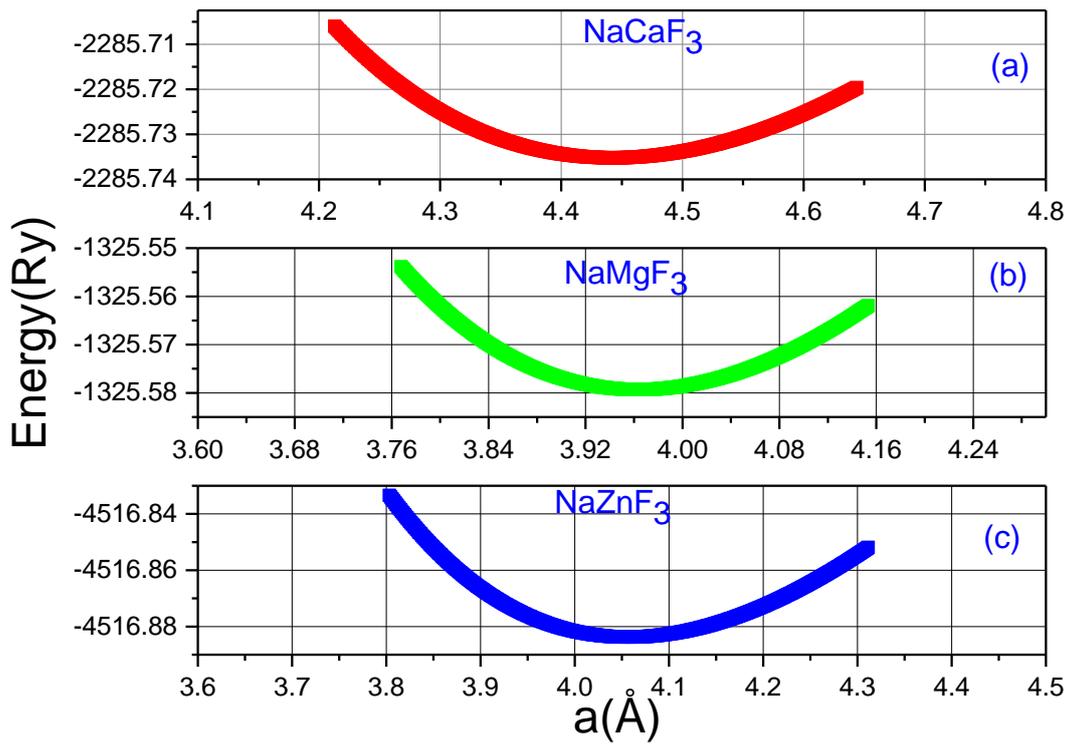

*Fig. 2: The total energy of the Cubic Fluoro-perovskites NaBF$_3$ (B= Ca, Mg or Zn) materials as a function of the lattice parameter a (Å), using the GGA-PBE.*

**III.2 Electronic properties of the cubic Fluoro-perovskites NaBF3 (B= Ca, Mg or Zn)**

To compare the electronic properties of the cubic Fluoro-perovskites NaBF$_3$ (B= Ca, Mg or Zn) given in this work, with those found in the literature, we provide such a comparison in Table 2, for the present work and the existing values of the energy Bandgap of the Fluoro-perovskites NaBF$_3$ (B= Ca, Mg or Zn) materials. At first glance, the band gap values obtained by mBJ-GGA approximation are bigger than those found in the literature, for the three studied Fluoro-perovskites NaBF$_3$ (B= Ca, Mg or Zn) compound.

|                     | **NaCaF$_3$**                              | **NaMgF$_3$** | **NaZnF$_3$** |
|---------------------|--------------------------------------------|---------------|---------------|
| Present study       | 7.067 eV                                   | 7.206 eV      | 4.652 eV      |
| Existing literature | 4.773 eV [43]<br>5.02 eV [52]<br>5.115 eV [53] | 5.90 eV [54]  | 3.35 eV [26]  |

*Table 2: A comparison between the present work and the existing literature for energy Bandgap of the Fluoro-perovskites NaBF$_3$ (B= Ca, Mg or Zn) materials.*

On the other hand, the above-cited results are well described in detail in the figure 3. The different Figs. 3a, 3b and 3c illustrate the band structure, the total density of states and the partial density of states and the band structure of the solar perovskite NaZnF$_3$, respectively.

From Fig. 3a, it is clear from the band structure, that the compound NaZnF$_3$ is a semiconductor with the band gag value 4.652 eV. In addition, Fig. 3b shows that the Na-2p orbital is the most contributing one in the conduction band region. While the orbitals Ca-2p and Ca-4s are impacting only the valence band region of the total DOS. The contribution of the F-2p is noticeable only in the conduction band. Concerning the band structure of the solar perovskite

NaCaF$_3$, our results are presented in Fig. 3a. From this figure, it is shown that this material presents a semi-conductor character, confirming the Fig. 3c findings. In addition, we found a direct band gap at the point Γ.

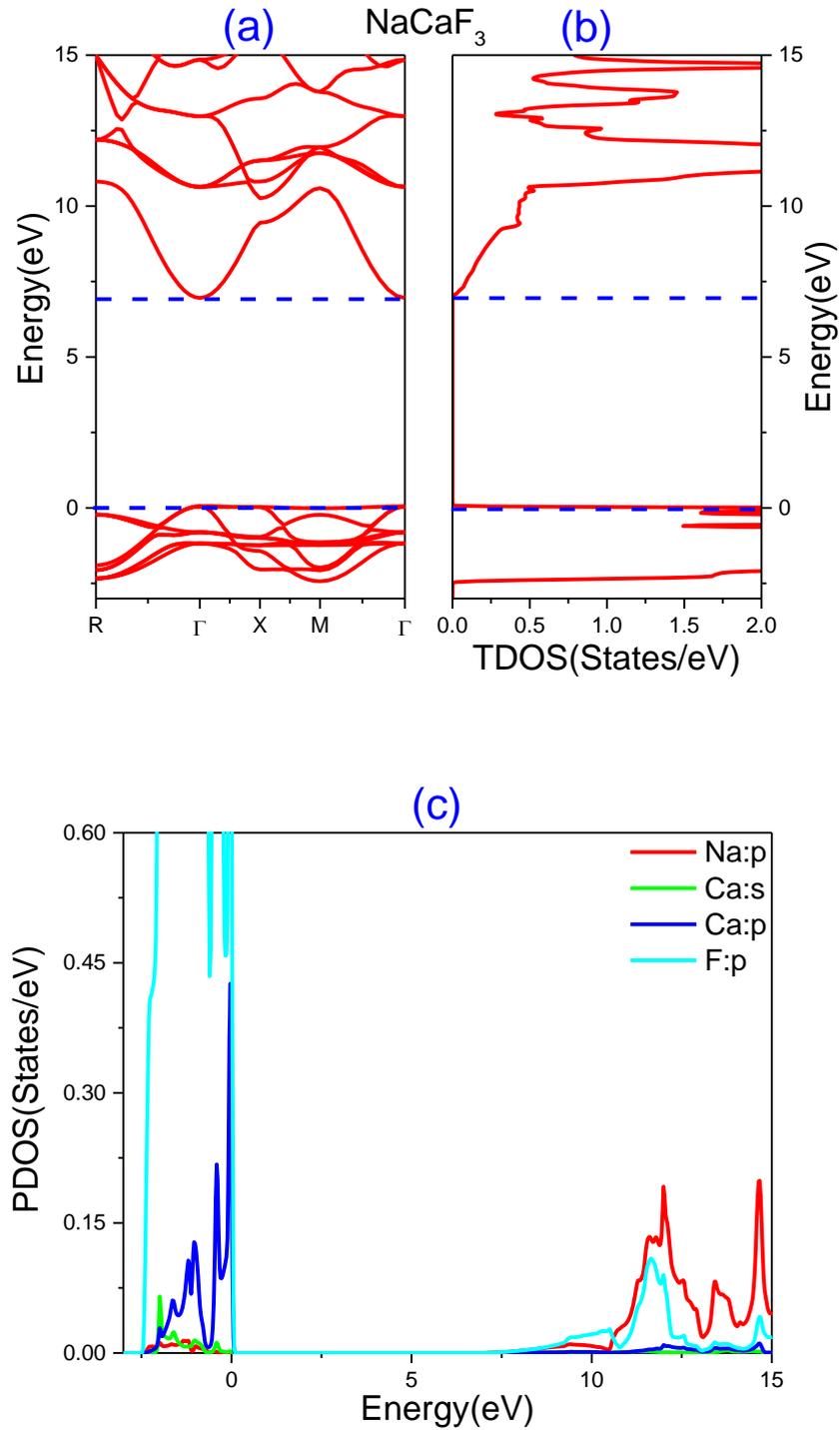

*Fig.3: Band structure (a), total density of states (b) and partial density of states (c) of the solar perovskite NaCaF$_3$ material.*

When considering the solar perovskite NaMgF$_3$, the obtained results are presented in Figs. 4a, 4b and 4c. The semiconductor character of the compound NaMgF$_3$ is well presented in Fig. 4a, with a band gap value of about 7.206 eV. From Fig. 4b, the orbitals Mg-2p and F-2p are contributing to both conduction and valence bands. While the effect of Mg-1s orbital is present only in the valence band. The partial and total density of states of the solar perovskite NaMgF$_3$ material presented in Fig. 4c, confirmed the semi-conductor behavior of this compound, as found in Fig. 4a. The band structure of the solar perovskite NaMgF$_3$ material is showed an indirect band gap between the points M and Γ, see Fig. 4a.

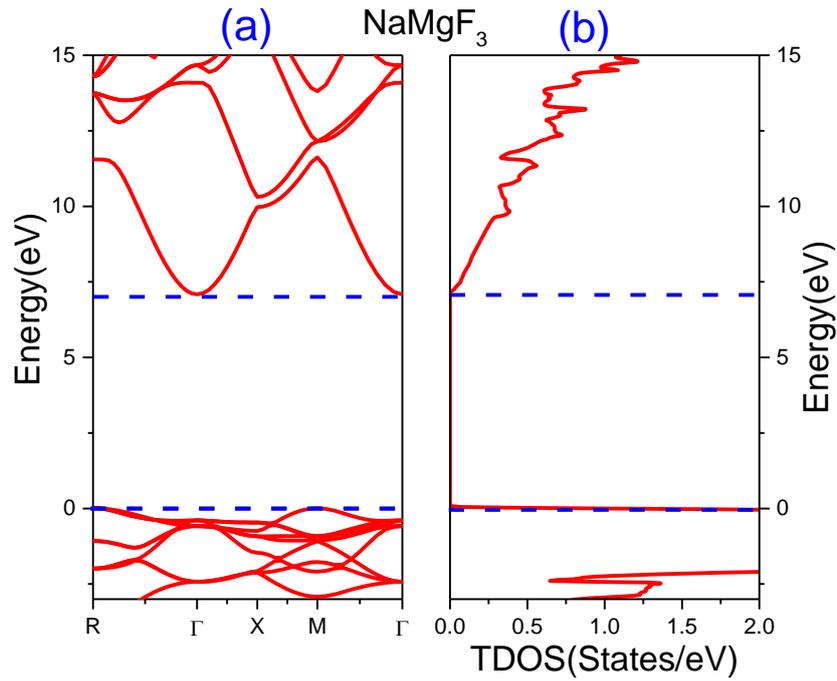

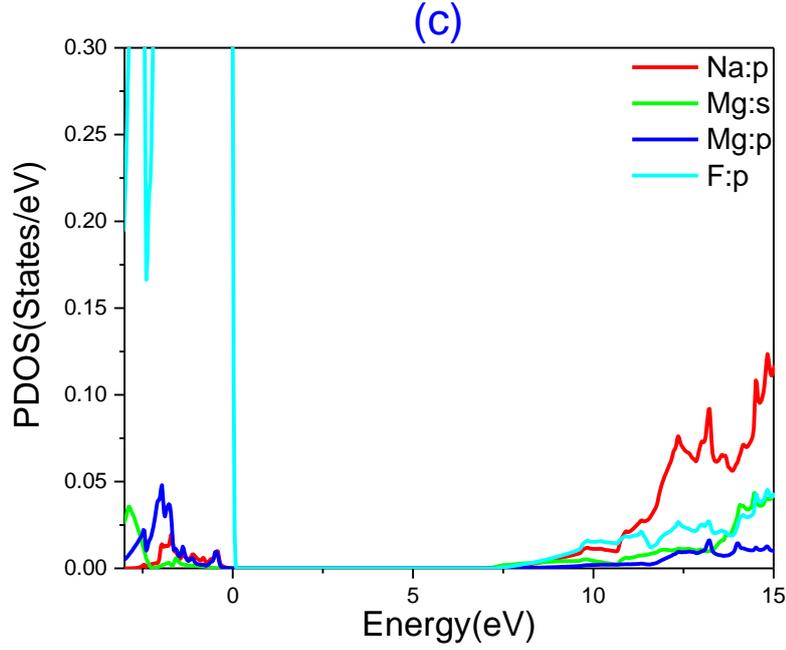

*Fig. 4: Band structure (a), total density of states (b) and partial density of states (c) of the solar perovskite NaMgF$_3$ material.*

On the other hand, when envisaging the band structure, the total and partial density of states of the solar perovskite NaZnF$_3$ material, our results are summarized in Figs. 5a, 5b and 5c, respectively. According to Figs. 5a and 5b, the NaZnF$_3$ compound exhibits an intrinsic semi-conductor character with a bang gap value of 4.652 eV. The partial density of states, plotted in Fig. 5b, shows that the orbital Zn-3d is acting mainly in the valence band region of the total DOS. While the orbitals Zn-2p and F-2p are mainly affecting the conduction band region. The presence of an indirect band gap has been found for the NaZnF$_3$ compound, between the points M and Γ, as it is illustrated in Fig. 5a.

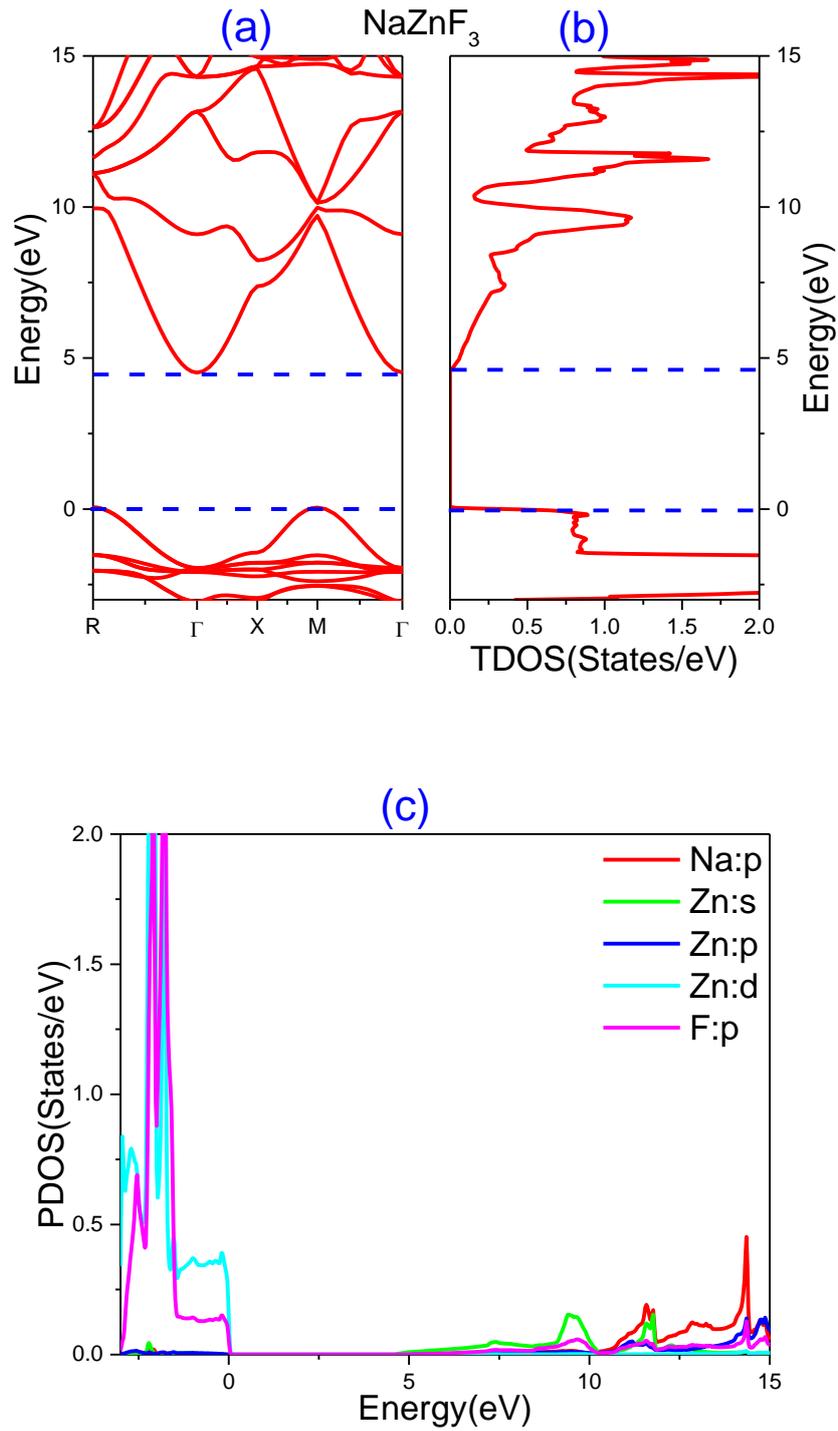

*Fig.5: Band structure (a), total density of states (b) and partial density of states (c) of the solar perovskite NaZnF$_3$ material.*

## IV. Time-Dependent Density Functional Theory method (TDDFT)

The time-dependent DFT (TDDFT) method was used to study the optical properties in the excited state, such as optical absorption. In this section, we adopt real space and real-time TDDFT.

The absorption spectrum and linear response of photo-generated carriers in the external field were described by using the time-dependent Kohn-Sham equation [54]:

$$i\frac{\partial \phi(t)}{\partial t} = [Ec + V_{KS}(t)]\phi(t) \qquad (2)$$

where $E_C$ stands for the kinetic operator; $V_{XC}$ is the exchange and correlation potential.

The optical properties of perovskite CsPbI$_3$ material were studied using the dielectric function ε(ω) that describes the system response to an external electromagnetic field through the interaction of photons with electrons [55]:

$$\varepsilon(\omega) = \varepsilon_1(\omega) + i\varepsilon_2(\omega) \qquad (3)$$

where $\varepsilon_1(\omega)$ and $\varepsilon_2(\omega)$ are the real and imaginary parts of the dielectric function, respectively. The real part of the dielectric function $\varepsilon_1(\omega)$ means the dispersion of the incident photons by the material. It can be reached by using the Kramer– Kronig relationship [56].

$$\varepsilon_1(\omega) = 1 + \frac{2}{\pi}\int_0^\infty \frac{\varepsilon_2(\omega')\omega' d\omega'}{\omega'^2 - \omega^2} \qquad (4)$$

While the imaginary part $\varepsilon_2(\omega)$ is directly associated to the electronic band structure and represents the light absorption in the crystal. It is given by the following equation:

$$\varepsilon_2(\omega) = \frac{4\pi^2 e^2}{m^2\omega^2}\sum_{i,j}|<i|M|j>|^2 \times (f_i(1-f_i))\delta(E_f - E_i - \hbar\omega)d^3k \qquad (5)$$

## V. Optical properties of the three solar fluoro-perovskites NaBF$_3$ (B= Ca, Mg or Zn)

In this part, we apply the mBJ-GGA approximation to illustrate the corresponding results. In fact, the Fig. 6 illustrates the real and imaginary parts of the dielectric tensor spectrum of the perovskite NaCaF$_3$ material. The same parameters have been examined by this method for the perovskites NaMgF$_3$ and NaZnF$_3$, in Figs. 7 and 8, respectively.

From these figures, it is found that the real part of the dielectric function plots increased with the increasing photon energy from the static values to reach maximum values of about 120 at 0.4 eV for the compound NaCaF$_3$, 280 at 0.2 eV for the material NaMgF$_3$ and 150 at 0.3 eV for the perovskite NaZnF$_3$.

The negative values of the real part of the dielectric function correspond to the metallic behavior of the three perovskites NaBF$_3$ (B=Ca, Mg or Zn). In fact, the incident electromagnetic waves undergo reflections in these parts of the energy. The second maxima reveal the presence of three transition frequencies. The imaginary part of the dielectric function starts from 0.25 eV for NaCaF$_3$ compound, 0.35 eV for NaMgF$_3$ material and 0.22 eV for NaZnF$_3$ perovskite. In addition, the imaginary part of the dielectric function is related to the absorption of radiation by the material. Also, the origin of the spectrum peaks was based on the density of states.

.

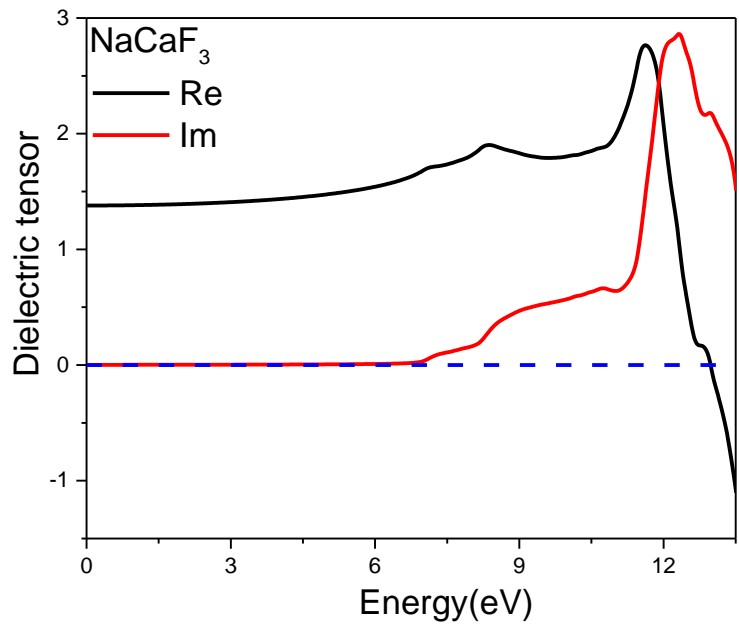

*Fig.6: The real and imaginary parts of the dielectric tensor spectrum using mBJ-GGA approximation of the perovskite NaCaF$_3$ material.*

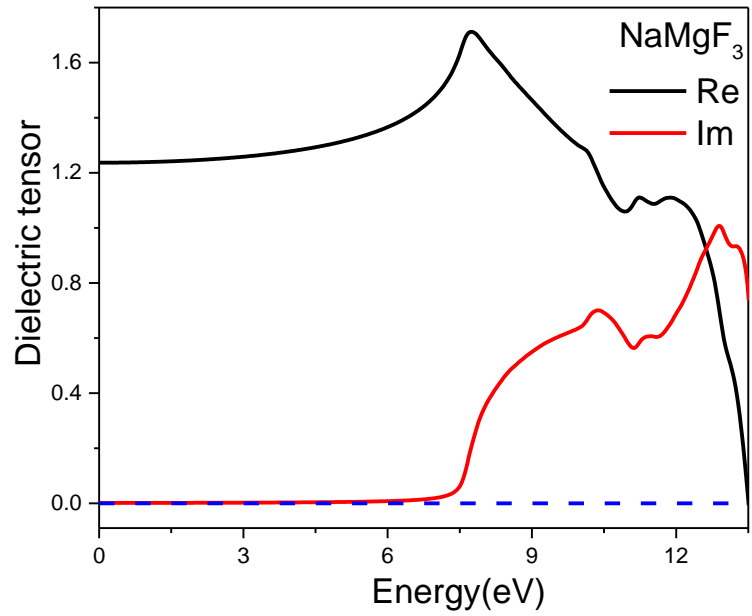

*Fig.7: The real and imaginary parts of the dielectric tensor spectrum using mBJ-GGA approximation of the perovskite NaMgF$_3$ material.*

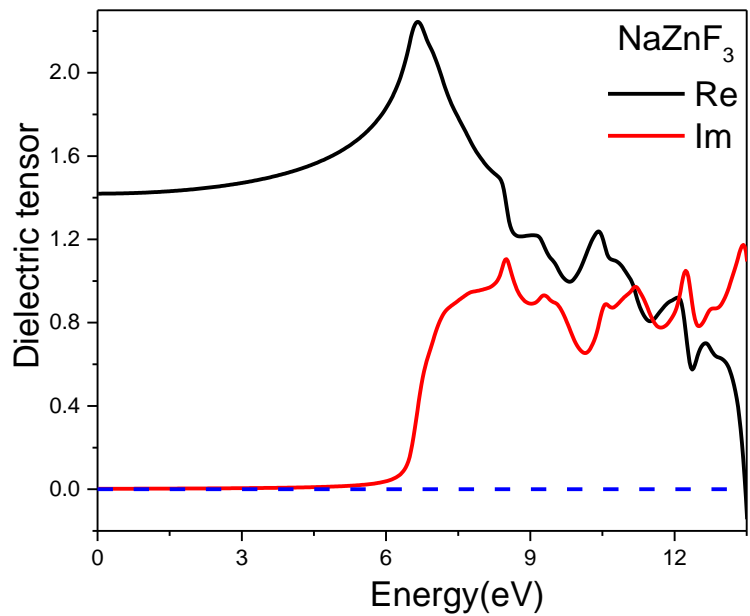

*Fig.8: The real and imaginary parts of the dielectric tensor spectrum using mBJ-GGA approximation of the perovskite NaZnF$_3$ material.*

## VI. Conclusion

In this paper, we have used the Quantum Espresso code under the pseudo-potentials wave method to inspect the structural, electronic and optical properties of the Sodium Based Cubic Fluoro-perovskites: NaBF$_3$ (B= Ca, Mg or Zn). Such simulation method is based on the density functional theory (DFT) and (TDDFT) methods using the mBJ-GGA approximation.

Firstly, we have considered the optimized unit cells for these compounds screening that the sodium-based cubic fluoro-perovskite NaZnF$_3$ perovskite is the more stable one. Also, we found that all these compounds exhibit a semi-conductor character with the band gag values: 4.50 eV, 4.00 eV and 3.20 eV for NaCaF$_3$, NaMgF$_3$ and NaZnF$_3$, respectively. In particular, the perovskite NaZnF$_3$ compound exhibits an intrinsic semi-conductor behavior. Besides, the solar perovskite NaCaF$_3$ showed a direct band gap at the point Γ, while the compounds NaMgF$_3$ and NaZnF$_3$ showed an indirect band gap between the points M and Γ. When examining the total and partial density of states of the three solar perovskites NaBF$_3$ (B=Ca, Mg or Zn) material, our results reveal that the Na-2p orbital is the most contributing one in the conduction band region for the perovskite NaCaF$_3$. While the orbitals Ca-2p and Ca-4s are influencing mainly the valence band region of the total DOS. The contribution of the F-2p is noticeable only in the conduction band for this compound. For the perovskite NaMgF$_3$, the orbitals Mg-2p and F-2p are mainly contributing to both the conduction and the valence bands. While, the effect of Mg-1s orbital is present only in the valence band. Finally, the orbital Zn-3d is acting mainly in the valence band region of the total DOS for the NaZnF$_3$ compound. While, the orbitals Zn-2p and F-2p are mainly affecting the conduction band region.

**References:**

[1] J. Fayos and J. Tornero, Crystallographic phase transition of NH$_4$ZnF$_3$ by single crystal XRD, *Ferroelectrics Letters Section,* 1993, **16**, 43, http://dx.doi.org/10.1080/07315179308204258.

[2] M. Hidaka, Z. Y. Zhou, and S. Yamashita, Structural phase transitions in KCdF$_3$ and K$_{0.5}$Rb$_{0.5}$CdF$_3$, *Phase Trans*. 1990, **20**, 83, http://dx.doi.org/10.1080/01411599008206869

[3] Y. Zhao et al., Critical phenomena and phase transition of perovskite — data for NaMgF$_3$ perovskite. Part II, *Phys. Earth Planet. Inter*. 1993, **76**, 17, http://dx.doi.org/10.1016/0031-9201(93)90051-A.

[4] F. Koussinssa and M. Diot, Propriétés thermodynamiques des trifluorocadmiates de rubidium et de césium (RbCdF$_3$, CsCdF$_3$) de 10 à 300 K: Transition de phase dans les fluoroperovskites Thermodynamic properties of rubidium and cesium trifluorocadmiates (RbCaF$_3$, CsCdF$_3$) from 10 to 300 K: Phase transitions in fluoroperovskites, *Thermochim. Acta,* 1993*,* **216**, 87, http://dx.doi.org/10.1016/0040-6031(93)80382-K.

[5] J. W. Flocken, R. A. Guenther, J. R. Hardy, and L. L. Boyer, A Priori Predictions of Phase Transitions in KCaF$_3$ and RbCaF$_3$: Existence of a New Ground State, *Phys. Rev. Lett.* 1986, **56**, 1738, http://dx.doi.org/10.1103/PhysRevLett.56.1738.

[6] U. J. Cox, A. Gibaud, and R. A. Cowley, Effect of Impurities on the First-Order Phase Transition of KMnF$_3$, *Phys. Rev. Lett*. 1988, **61**, 982, http://dx.doi.org/10.1103/PhysRevLett.61.982.

[7] A. Ratuszna, Structural phase transitions in (K$_{1-x}$Na$_x$)MnF$_3$ and (K$_{1-x}$Li$_x$)MnF$_3$ mixed perovskite crystals, *J. Phys. Condens. Matter*, 1993, **5**, 841, http://dx.doi.org/10.1088/0953-8984/5/7/010.

[8] S. Idrissi, H. Labrim, L. Bahmad and A. Benyoussef, Structural, Electronic and Magnetic Properties of the Rare Earth-Based Solar Perovskites: GdAlO$_3$, DyAlO$_3$ and HoAlO$_3$", *Journal of Superconductivity and Novel Magnetism*, 2021, **34**, 2371, http://dx.doi.org/10.1007/s10948-021-05900-3.

[9] S.A. Wolf, D.D. Awschalom, R.A. Buhrman, J.M. Daughton, S. von Molnar, M.L. Roukes, A.Y. Chtchelkanova and D.M. Treger, Spintronics: A spin-based electronics vision for the future, *Science* 2001, **294**, 1488, http://dx.doi.org/10.1126/science.1065389.

[11] S. Idrissi, H. Labrim, L. Bahmad, A. Benyoussef, DFT and TDDFT studies of the new inorganic perovskite CsPbI$_3$ for solar cell applications, *Chemical Physics Letters*, 2021, **766**, 138347, https://doi.org/10.1016/j.cplett.2021.138347.

[12] H. Ullah, S. Naeem, G. Murtaza, R. Khenata, M.N. Khalid, First principle study of CsSrM$_3$ (M = F, Cl). Phys B, 2013, **414,** 91, https://doi.org/10.1016/j.physb.2013.01.009.

[13] H. Ullah, G. Murtaza, R. Khenata, S. Muhammad, A.H. Reshak, K.M. Wong, S.B. Omran, Z.A. Ahmed, Structural, chemical bonding, electronic and magnetic properties of KMF$_3$ (M = Mn, Fe Co, Ni) compounds. *Computational Materials Science, 2014,* **85,** 402, http://dx.doi.org/10.1016/j.commatsci.2013.12.054.

[14] H. Ullah, G. Murtaza, R. Khenata, S. Mohammad, S. Naeem, M.N. Khalid, A. Manzar, Structural, elastic, electronic and optical properties of CsMCl$_3$ (M=Zn, Cd), *Phys B*, 2013, **420**, 15, http://dx.doi.org/10.1016/j.physb.2013.03.011

[15] A. Guzik, E. Talik, A. Pajaczkowska, S. Turczynski, J. Kusz J, Magnetic properties of manganese doped PrAlO$_3$ monocrystalline fibres, *Mater Sci Poland,* 2014, **32,** 633, http://dx.doi.org/10.2478/s13536-014-0240-y.

[16] A. M. Glazer, Simple ways of determining perovskite structures, *Acta Cryst.* 1975, **A31**, 756, http://dx.doi.org/10.1107/S0567739475001635.

[17] F. A. Kasan-Ogly and V. E. Naish, The immanent chaotization of crystal structures and the resulting diffuse scattering. IV. Diffuse scattering in perovskites with two-dimensional movable objects (tilting), *Acta Cryst.* 1986, **B42**, 325, http://dx.doi.org/10.1107/S0108768186098154.

[18] K. Umemoto, R.M. Wentzcovitch, D.J. Weidner, J.B. Parise, NaMgF3: A low-pressure analog of MgSiO$_3$, *Geophys. Res. Lett*. 2006, **33,** L15304, http://dx.doi.org/10.1029/2006GL026348.

[19] J. H. Liu, C.D. Martin, J.B. Parise, D.J. Weidner, Crystal chemistry of NaMgF$_3$ perovskite at high pressure and temperature, *Am. Mineral.* 2005, **90** 1534, http://dx.doi.org/10.2138/am.2005.1708.

[22] K. Umemoto, R.M. Wentzcovitch, Potential ultrahigh pressure polymorphs of ABX$_3$-type compounds, *Phys. Rev. B*, 2006, **74,** 224105, http://dx.doi.org/10.1103/PhysRevB.74.224105.

[23] T.F. Soules, E.J. Kelly, D.M. Vaught, J.W. Richardson, Energy-Band Structure of SrTiO$_3$ from a Self-Consistent-Field Tight-Binding Calculation, *Phys. Rev. B*, 1972, **6**, 1519, http://dx.doi.org/10.1103/PhysRevB.6.1519.

[24] R.G. Shulman, Y. Yafet, P. Eisenberger, W.E. Blumberg, Observations and interpretation of x-ray absorption edges in iron compounds and proteins, *Proc. Nat. Acad. Sci. U. S.* 1976, **A73,** 1384; https://doi.org/10.1073/pnas.73.5.138.

[25] M. Akaogi, Y. Shirako, H. Kojitani, T. Nagakari, H. Yusa, K. Yamaura, High-pressure transitions in NaZnF$_3$ and NaMnF$_3$ perovskites, and crystal-chemical characteristics of perovskite-postperovskite transitions in ABX$_3$ fluorides and oxides, *Physics of the Earth and Planetary Interiors,* 2014, **228**, 160, http://dx.doi.org/ 10.1016/j.pepi.2013.09.001.

[26] R. Arar, T. Ouahrani, D. Varshney, R. Khenata, G. Murtaza, D. Rached, A. Bouhemadou, Y. Al-Douri, S. Bin Omran, A.H. Reshak, Structural, mechanical and electronic properties of sodium based Fluoro-perovskites NaBF$_3$ (B=Mg, Zn) from first-principle calculations, *Materials Science in Semiconductor Processing*, 2015, **33**, 127, https://doi.org/10.1016/j.mssp.2015.01.040.

[27] S. Idrissi, S. Ziti, H. Labrim, L. Bahmad, Band gaps of the solar perovskites photovoltaic CsXCl3 (X=Sn, Pb or Ge), *Materials Science in Semiconductor Processing*, 2021, **122**, 105484, http://dx.doi.org/10.1016/j.mssp.2020.105484.

[28] S. Idrissi, O. Mounkachi, L. Bahmad, A. Benyoussef, Study of the electronic and opto-electronic properties of the perovskite KPbBr$_3$ by DFT and TDDFT methods, *Computational Condensed Matter*, 2021, **33**, e00617, https://doi.org/10.1016/j.cocom.2021.e00617.

[29] S. Idrissi, H. Labrim, L. Bahmad, A. Benyoussef, Study of the solar perovskite CsMBr$_3$ (M=Pb or Ge) photovoltaic materials: Band-gap engineering, *Solid State Sciences*, 2021, **118**, 106679, https://doi.org/10.1016/j.solidstatesciences.2021.106679.

[30] P. Giannouzzi et al., Quantum Espresso: A Modular and Open-Source Software Project for Quantum Simulations of Materials, *Journal of Physics: Condensed Matter*, 2009, **21,** https://doi.org/10.1088/0953-8984/21/39/395502.

[31] D. R. Hamann, M. Schlüter, C. Chiang, Norm-conserving pseudopotentials, *Phys. Rev. Lett.*, 1979, **43,** 1494, https://doi.org/10.1103/PhysRevLett.43.1494.

[32] J. P. Perdew, J. A. Chevary, S. H. Vosko, K. A. Jackson, M. R. Pederson, D. J. Singh, C. Fiolhais, Atoms, molecules, solids, and surfaces: applications of the generalized gradient approximation for exchange and correlation, *Phys. Rev. B*, 1992, **46**, 6671, http://dx.doi.org/10.1103/PhysRevB.46.6671.

[33] H. J. Monkhorst, J.D. Pack, Special points for Brillouin-zone integrations, *Phys. Rev. B*, 19976, **13**, 5188, http://dx.doi.org/10.1103/PhysRevB.13.5188.

[34] D.J. Head and M.C. Zerner, A Broyden–Fletcher–Goldfarb–Shanno optimization procedure for molecular geometries, *Chemical Physical Letters*, 1985, *122*, 264, http://dx.doi.org/10.1016/0009-2614(85)80574-1.

[35] S. Idrissi, S. Ziti, H. Labrim, L. Bahmad, Half-Metallicity and Magnetism in the Full Heusler Alloy $Fe_2MnSn$ with $L2_1$ and XA Stability Ordering Phases, *J Low Temp Phys.*, 2021, **202**, 343, https://doi.org/10.1007/s10909-021-02562-2

[36] S. Idrissi, S. Ziti, H. Labrim, L. Bahmad, Sulfur doping effect on the electronic properties of zirconium dioxide $ZrO_2$, *Materials Science and Engineering: B*, 2021, **270**, 115200, https://doi.org/10.1016/j.mseb.2021.115200.

[37] S. Idrissi, H. Labrim, S. Ziti, L. Bahmad, A DFT study of the equiatomic quaternary Heusler alloys ZnCdXMn (X=Pd, Ni or Pt), *Solid State Communications*, 2021, **331**, 114292, https://doi.org/10.1016/j.ssc.2021.114292.

[38] S. Idrissi, H. Labrim, S. Ziti And L. Bahmad , Investigation of the physical properties of the equiatomic quaternary Heusler alloys CoYCrZ (Z= Si and Ge) : A DFT study, *journal of applied physics A*, 2020, **126**, 190, http://dx.doi.org/10.1007/s00339-020-3354-6.

[39] S. Idrissi, H. Labrim, S. Ziti, L. Bahmad, Structural, electronic, magnetic properties and critical behavior of the equiatomic quaternary Heusler alloy CoFeTiSn, *Physics Letters A*, 2020, **384**, 126453, https://doi.org/10.1016/j.physleta.2020.126453.

[40] S. Idrissi, H. Labrim, S. Ziti, L. Bahmad, Critical magnetic behavior of the Rare Earth Based Alloy GdN: Monte Carlo simulations and DFT method, Journal of Materials Engineering and Performance, 2020, **29**, 7361, https://doi.org/10.1007/s11665-020-05214-w.

[41] R. Burriel, et al., $KZnF_3$ cubic perovskite. Heat capacity and lattice dynamics, *J. Phys. C. Solid State Phys*. 1987, **20**, 2819e2827 ; http://dx.doi.org/10.1088/0022-3719/20/19/011.

[42] K. Momma and F. Izumi, VESTA 3 for three-dimensional visualization of crystal, volumetric and morphology data, *J. Appl. Crystallogr.*, 2011, **44**, 1272, http://dx.doi.org/10.1107/S0021889811038970.

[43] S.S.A. GILLANI, F. Nisar, M. SHAKIL, et al. Ultra-wide bandgap semiconductor behavior of $NaCaF_3$ fluoro-perovskite with external static isotropic pressure and its impact on optical properties: First-principles computation, 2021, https://doi.org/10.21203/rs.3.rs-336152/v1.

[44] H. Bouafia, B. Sahli, S. Hiadsi, B. Abidri, D. Rached, A. Akriche, M. N. Mesli, Theoretical investigation of structural, elastic, electronic, and thermal properties of KCaF$_3$, K$_{0.5}$Na$_{0.5}$CaF$_3$ and NaCaF$_3$ perovskites, *Superlatt. Microstruct.* 2015 **82**, 525, http://dx.doi.org/10.1016/j.spmi.2015.03.004.

[45] M. J. Mehl, Pressure dependence of the elastic moduli in aluminum-rich Al-Li compounds, *Phys. Rev. B*, 1993, **47**, 2493, http://dx.doi.org/10.1103/PhysRevB.47.2493.

[46] M. Harmel, H. Khachai, M. Ameri, R. Khenata, N. Baki, A. Haddou, B. Abbar, S¸Ugur, S. Bin Omran, F. Soyalp, DFT-based ab initio study of the electronic and optical properties of cesium based fluoro-perovskite CsMF$_3$ (M = Ca and Sr), *Int. J. Modern Physics B*, 2012, **26**, 12501991; http://dx.doi.org/10.1142/S0217979212501998.

[47] M. H. Benkabou, M. Harmel, A. Haddou, A. Yakoubi, N. Baki, R. Ahmed, Y. Al-Douri, S.V. Syrotyuk, H. Khachai, R. Khenata, C. H. Voon, M. R. Johan, Structural, electronic, optical and thermodynamic investigations of NaBF$_3$ (X = Ca and Sr): First-principles calculations, *Chinese Journal of Physics*, 2018, **56,** 131, http://dx.doi.org/10.1016/j.cjph.2017.12.008.

[48] G. A. Geguzina, V.P. Sakhnenko, Correlation between the lattice parameters of crystals with perovskite structure, *Cryst. Rep.* 2004, **49,** 15, http://dx.doi.org/10.1134/1.1643959

[49] E.C.T. CHAO, H.T. EVANS JR, B.J. SKINNER, et al. Neighborite, NaMgF$_3$, a new mineral from the green river formation, South Ouray, Utah, *American Mineralogist: Journal of Earth and Planetary Materials*, 1961, **46**, 379.

[50] U. Yasutomo, K. Sadaomi, O. Tomoyuki, T. Kazuyoshi and N. Eiji, Structure Evaluation of Bio-Compatible Lead-Free Piezoelectric Materials by Crystal System Distinction and First Principles Calculations, *Nihon Kikai Gakkai Ronbunshu, A Hen*, 2006, **72**, 1472, https://doi.org/10.1299/kikaia.72.1472.

[51] D. A. C. Garcia-Castro, N. A. Spaldin, A. H. Romero, and E. Bousquet, Geometric ferroelectricity in fluoroperovskites, *Phys. Rev. B*, 2014, **89**, 104107, http://dx.doi.org/10.1103/PhysRevB.89.104107.

[52] K. Yang, Y. He, Y. Cheng, L. Che and L. Yao L, Ion-number-density-dependent effects on hyperfine transition of trapped $^{199}$Hg$^+$ ions in quadrupole linear ion traps, *Phys. Lett. A*, 2017, **381**, 890, http://dx.doi.org/10.1016/j.physleta.2017.02.006.

[53] Z.L. Li, X.Y. An, X.L. Cheng, X.M. Wang, H. Zhang, L.P. Peng and W.D. Wu, First-principles study of the electronic structure and optical properties of cubic Perovskite NaMgF$_3$, *Chinese Physics B,* 2014, **23**, 037104, http://dx.doi.org/10.1088/1674-1056/23/3/037104.

[54] K. Yabana, G. F. Bertsch, Time-dependent local-density approximation in real time, *Phys. Rev. B*, 1996, **54**, 4484, http://dx.doi.org/10.1103/PhysRevB.54.4484.

[55] J. Qiao, X. Kong, Z. X. Hu, F. Yang, W. Ji, High-mobility transport anisotropy and linear dichroism in few-layer black phosphorus, *Nat. Commun.* 2014, **5**, 4475, https://doi.org/10.1038/ncomms5475.

[56] D.C. Hutchings, M. Sheik-Bahae, D.J. Hagan, *et al.* Kramers-Krönig relations in nonlinear optics. *Opt Quant Electron*, 1992, **24**, 1, https://doi.org/10.1007/BF01234275.

**Figures:**

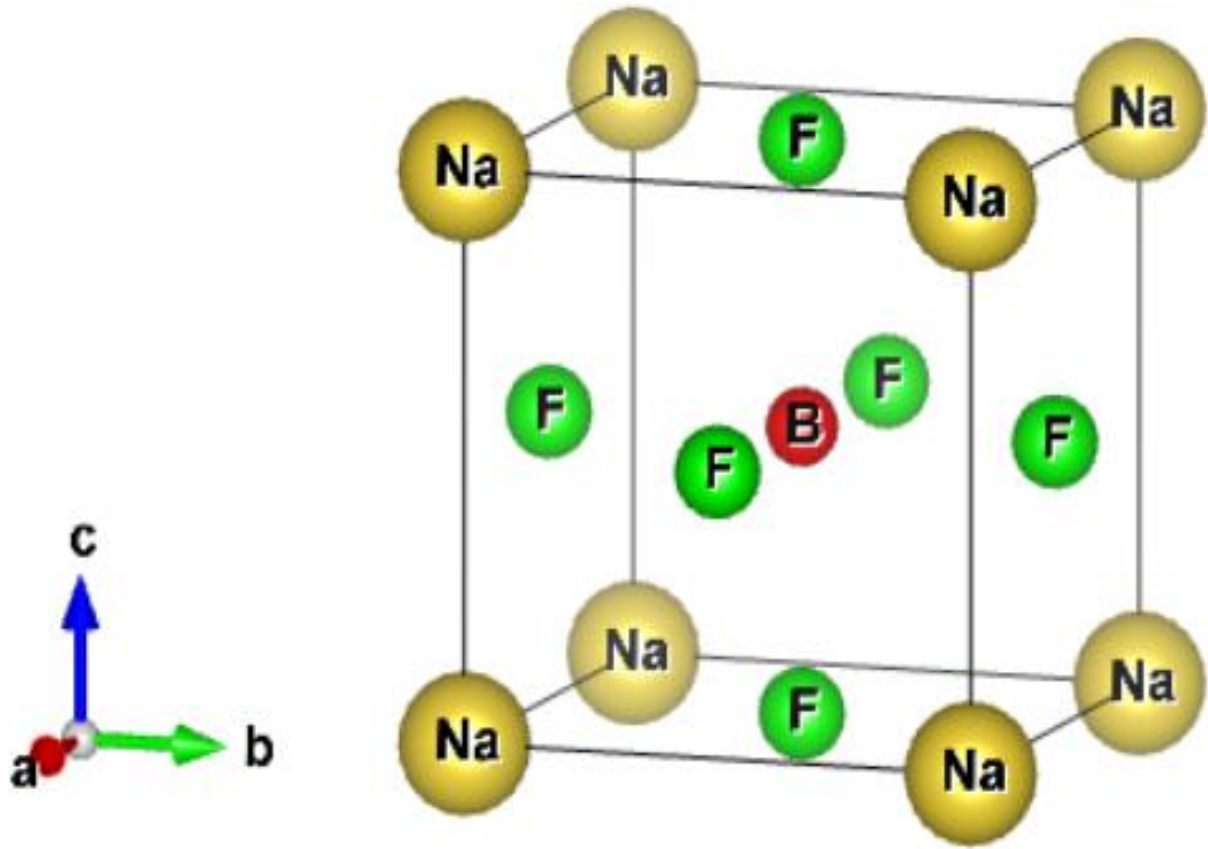

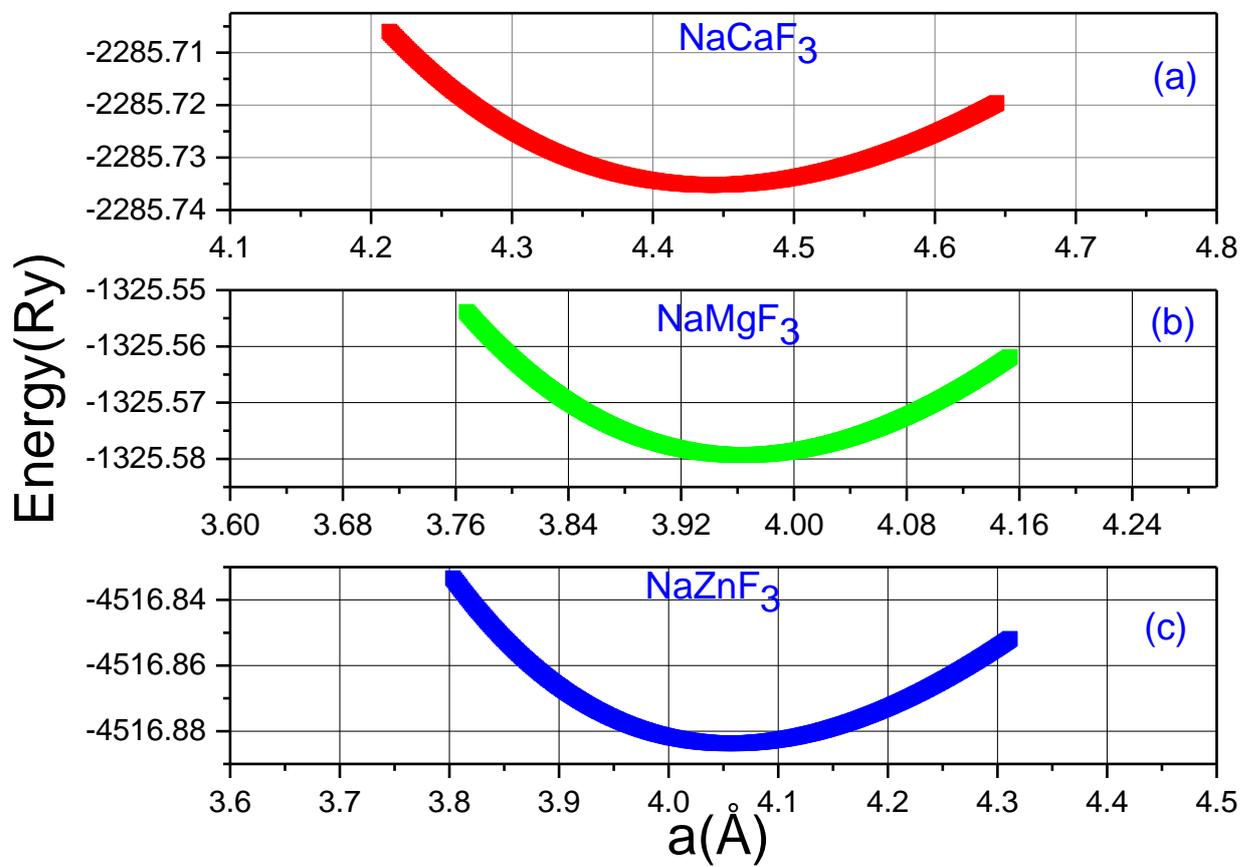

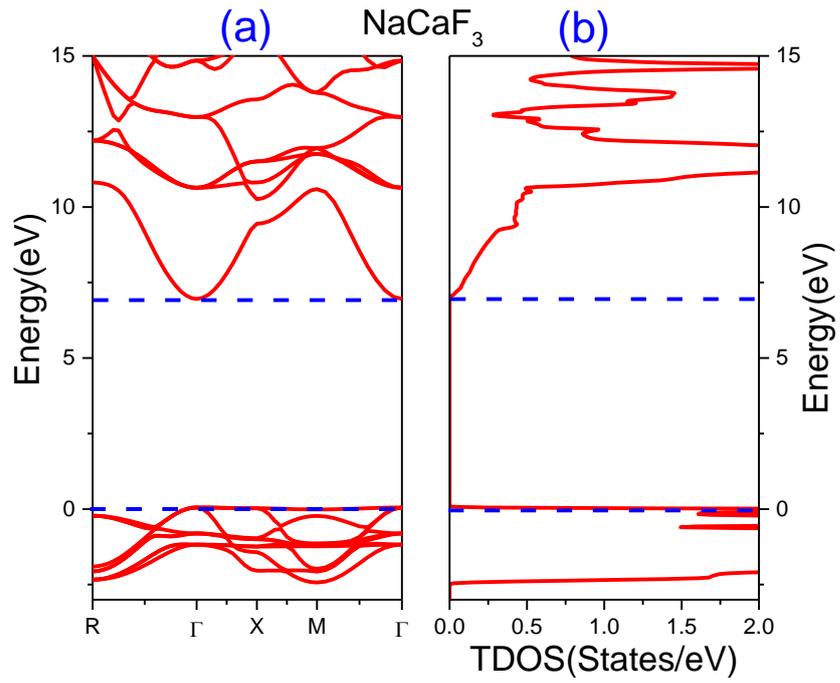

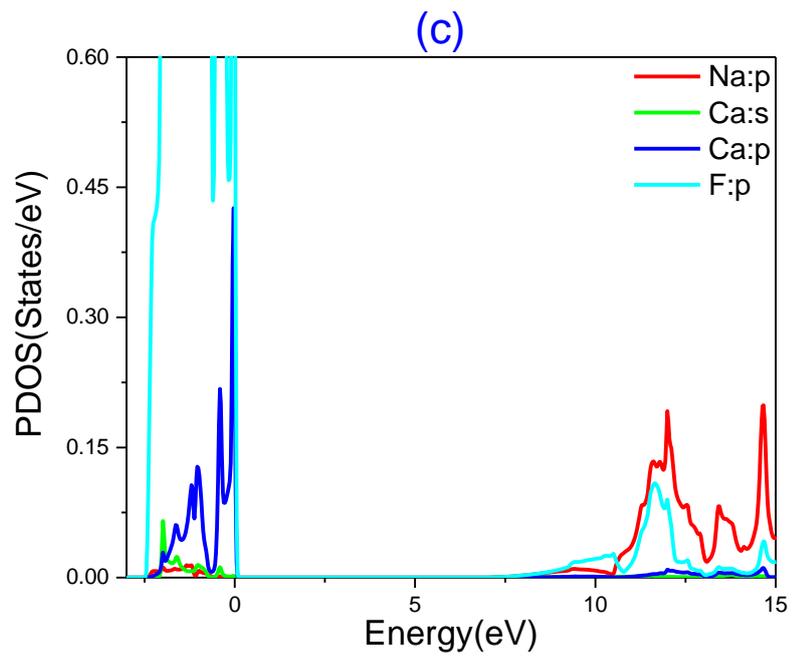

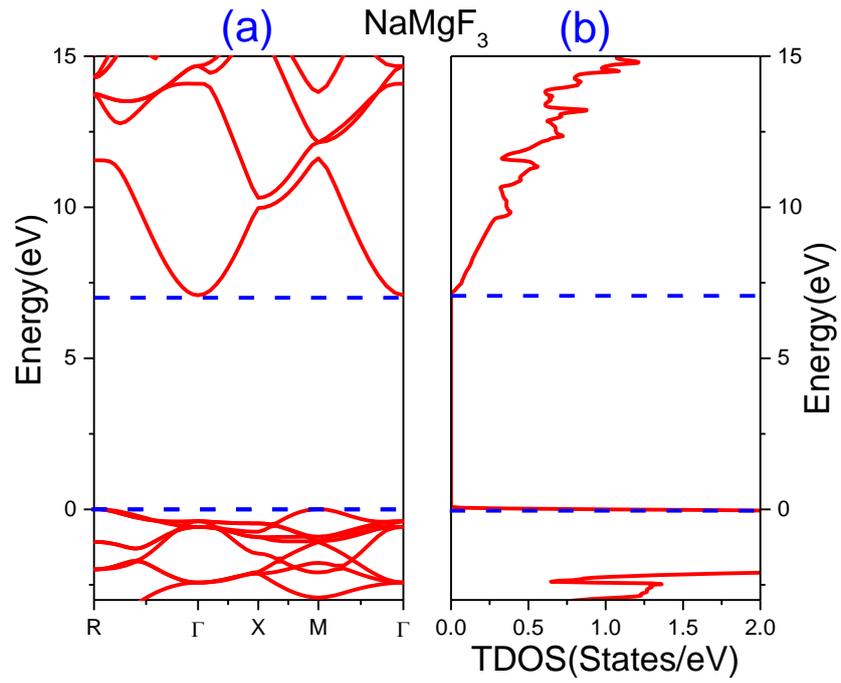
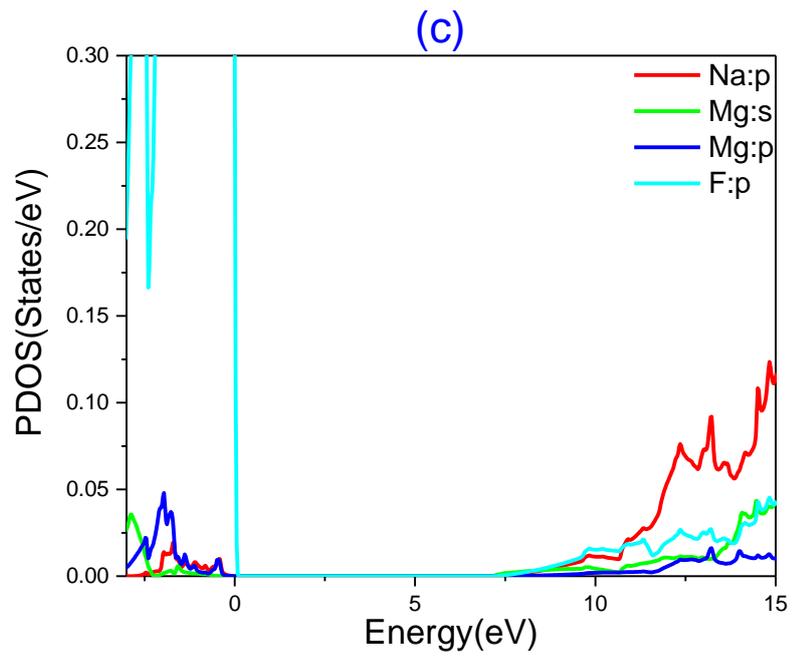

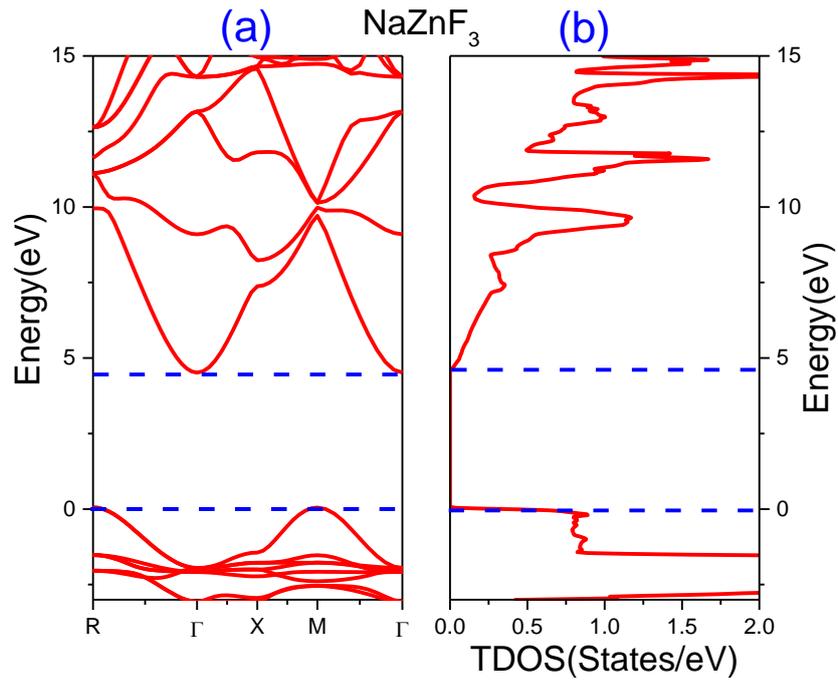

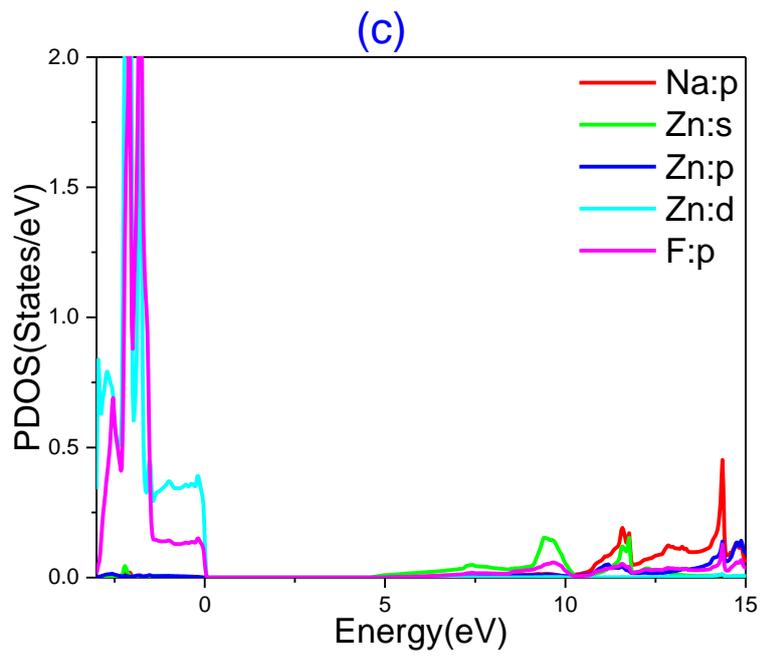

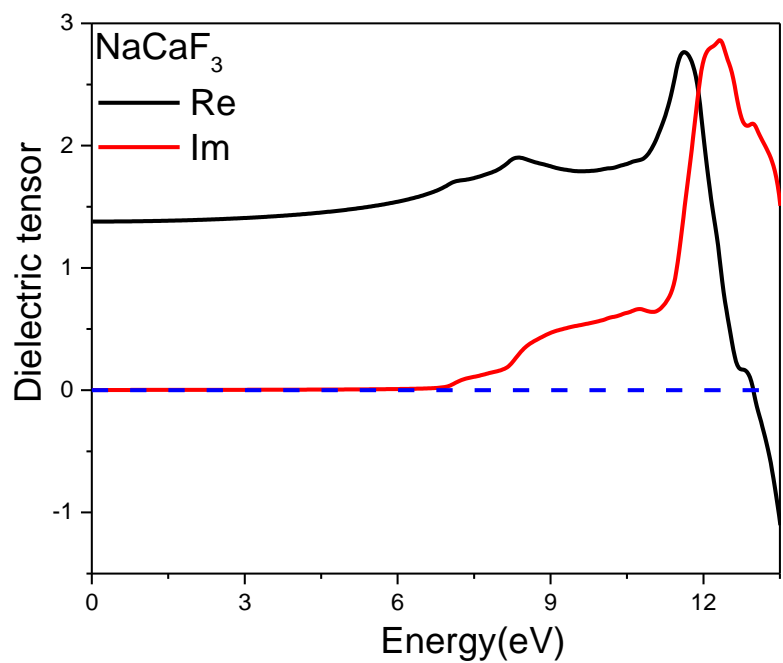

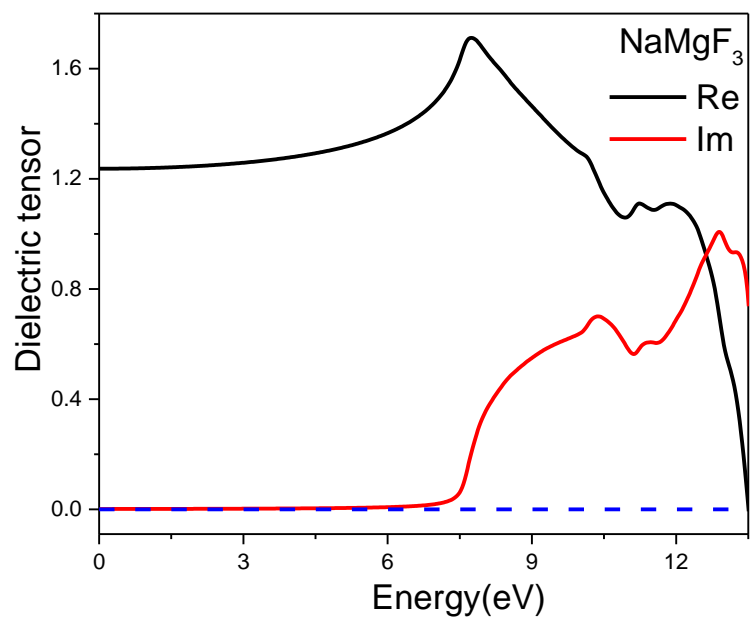

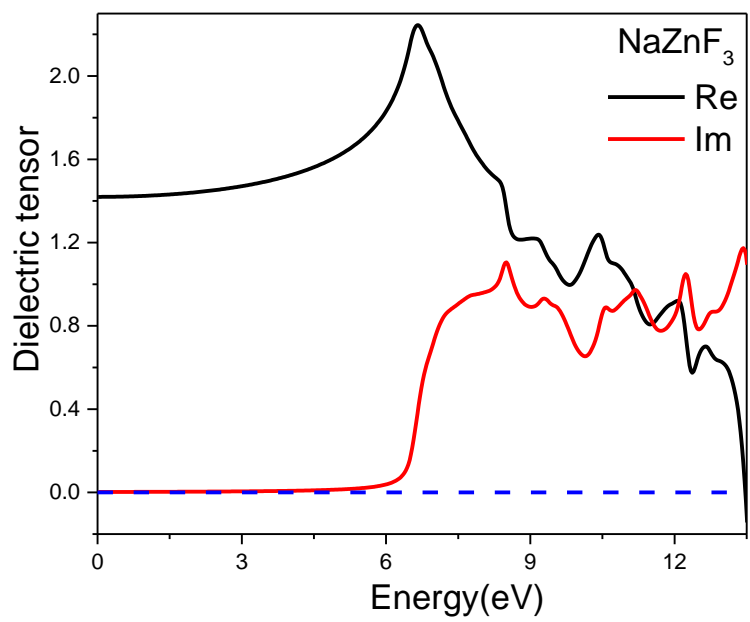